\documentclass[twocolumn,showpacs,preprintnumbers,amsmath,amssymb]{revtex4}
\usepackage{graphicx}
\usepackage{dcolumn}

\def\prl#1#2#3{{ Phys.   Rev.   Lett.  } {\bf #1}, #2 (#3)}

\def\pre#1#2#3{Phys.   Rev.   E {\bf #1}, #2 (#3)}

\def\epl#1#2#3{Europhys. Lett. {\bf #1}, #2 (#3)}

\def\jsm#1#2#3#{J.Stat.Mech. {\bf#1}, #2 (#3)}

\def\noi{\noindent}
\def\bc{\begin{center}}
\def\ec{\end{center}}
 \newcommand{\bea}{\begin{equation}}
 \newcommand{\eea}{\end{equation}\noi}
 \newcommand{\ber}{\begin{eqnarray}}
 \newcommand{\eer}{\end{eqnarray}\noi}
\begin{document}
\title{Nonequilibrium Work distributions for a trapped Brownian particle in a time dependent magnetic field}
\author{Arnab Saha[1] and A.M.Jayannavar[2]}
\address{[1] S.N.Bose National Centre For Basic Sciences, JD-Block,
Sector-III, Salt Lake City, Kolkata-700098, India \\
[2] Institute Of Physics, Sachivalaya Marg, Bhubaneswar 751005, India} 

\date{\today}
\begin{abstract}
We study the dynamics of a trapped, charged Brownian particle in presence of a time dependent magnetic field. We calculate work distributions for different time dependent protocols. In our problem thermodynamic work  is related to variation of vector potential with time as opposed to the earlier studies where the work is related to time variation of the potentials which depends only on the coordinates of the particle. Using Jarzynski identity and Crook's equality we show that the free energy of the particle is independent of the magnetic field, thus complementing the Bohr-van Leeuwen theorem. We also show that our system exhibits a parametric resonance in certain parameter space.     
\end{abstract}
\maketitle

Equilibrium statistical mechanics provides us an elegant framework to explain properties of a broad variety of systems in equilibrium. Close to equilibrium  the linear response formalism is very successful in the form of fluctuation-dissipation theorem and Onsager's reciprocity relations.  But no such universal framework exists to study systems driven far away from equilibrium. Needless to say that the most processes that occur in nature are far from equilibrium. In recent years there has been considerable interest in the nonequilibrium statistical mechanics of small systems. This has led to discovery of several rigorous theorems, called fluctuation theorems (FT) and related equalities \cite{1,2,3,4,5,6,7,8,9,10,11} for systems far away from equilibrium. Some of these 
theorems have been verified experimentally \cite{12,13,14,15,16} on single nanosystems in physical environment where fluctuations play a dominent role.   
We will focus on the Jarzynski identity \cite{4} and Crook's equality \cite{5} which deal with systems which are initially in thermal equilibrium and are driven far away from equilibrium irreversibly. Jarzynski identity relates the free energy change($\Delta F$) of the system when it is driven out of equilibrium by perturbing its Hamiltonian ($H_{\lambda}$) by an externally controlled time dependent protocol $\lambda(t)$, to the thermodynamic work(W) done on the system, given by 
\bea
W=\int_0^{\tau}\dot \lambda \frac{\partial H_{\lambda}}{\partial \lambda}dt,
\eea
over a phase space trajectory. Here $\tau$ is the time through which the system is driven. Jarzynski identity is,
\bea
\left<e^{-\beta W}\right>=e^{-\beta\Delta F}.
\eea
Crook's equality relates the ratio of the work distributions in forward and backward ( time reversed ) paths through which the system evolves. This relation is given by,
\bea
\frac{P_f(W)}{P_b(-W)}=e^{\beta W_d},
\eea
where, $P_f$ and $P_b$ are the distributions of work along forward and backward paths respectively. Here, the dissipative work $W_d=W-W_r$ and $W_r$ is the reversible work which is same as the free energy difference ($\Delta F$) between the initial and the final states of the system when driven through a reversible, isothermal path. If the system is driven reversibly all along the path, the work distribution will be $\delta(W-\Delta F)$, $W_d=0$ and $P_f=P_b$. Thus, the above identities are trivially true for reversibly driven system. Jarzynski identity follows from equation (3). Crooks relation follows from a more general Crooks identity which relates ratio of work probabilities of forward path and that of the reverse path to the dissipative work expended along the forward trajectory.
\par    
In this brief report, we will study the applicability of Jarzynski identity and Crooks equality in case of velocity dependent as well as time dependent Lorentz force which is derivable from a generalised potential, $U=q(\phi-\bold A(t).\bold v)$. Here, $\bold A$ is time dependent vector potential, $\phi$ is scaler potential, $q$ is the charge of a particle and $\bold v$ is its velocity. Different time dependent protocols for magnetic fields are considered. Consequently, we find that, the free energy difference obtained using Jarzynski and Crooks equality complements Bohr-van Leeuwen theorem \cite{17,18,19} and thus we arrive at  an alternative proof of Bohr-van Leeuwen theorem. This theorem states that in case of classical systems the free energy is independent of magnetic field and hence the thoerem concludes absence of diamagnetism in classical thermodyanamical equilibrium systems. We finally also show that our system, in presence of ac magnetic field exhibits parametric resonance in certain parameter regime.
\par
In an earlier related work \cite{19,20} a charged particle dynamics in overdamped limit is studied in the presence of harmonic trap and static magnetic field. The work distribution have been obtained analytically for different protocols. It is shown that work distribution depends explicitly on the magnetic field but not the free energy difference ($\Delta F$).  

 The model Hamiltonian for our isolated system is,            

\bea
H= \frac{1}{2m}\left[ \left(p_x+\frac{qB(t)y}{2}\right)^2+\left(p_y-\frac{qB(t)x}{2}\right)^2\right]+\frac{1}{2}k(x^2+y^2),
\eea
where $k$ is the stiffness constant of harmonic confinement. The  magnetic field $B(t)$ is applied in the z direction. The $x$ and $y$ components of the vector potential, $A_x$ $A_y$ are given by $-\frac{qB(t)y}{2}$ and $\frac{qB(t)x}{2}$ respectively. We have chosen symmetric guage here. The particle-environment interaction is modeled via Langevin equation including inertia \cite{21}, namely,  
 
\bea
m\ddot x=\frac{q}{2}\left[y\dot B(t)+2\dot yB(t)\right]-kx-\Gamma \dot x+\eta_x(t),
\eea
\bea
m\ddot y=-\frac{q}{2}\left[x\dot B(t)+2\dot xB(t)\right]-ky-\Gamma \dot y+\eta_y(t),
\eea
where $\Gamma$ is the friction coefficient and $\eta_x$ and $\eta_y$ are the Gaussian white noise along $x$ and $y$ direction respectively. This thermal noise has the following properties,
\ber
<\eta_i>=0 ; <\eta_i(t)\eta_j(t^{\prime})>=\delta_{ij}2\Gamma k_BT\delta(t-t^{\prime}).
\eer
With the above prescription for the thermal noise, the system approaches a unique equilibrium state in the absence of time dependent potentials. Denoting the protocol  
$\lambda (t)=\frac{q}{2}B(t)$
the thermodynamic work done by external magnetic field on the system upto time $\tau$  is,
 
\bea
W=-\frac{q}{2}\int_0^\tau (x\dot y-y\dot x)\dot B(t)dt.
\eea
We will like to emphasize  that this thermodynamic work is related to the time variation of the vector potential and can be identified as time variation of magnetic potential $-\bold \mu.\bold B$, $W=-\int_0^\tau \bold \mu. \frac {d\bold B}{dt} dt$, where induced magnetic moment is $\frac{q}{2}(x\dot y-y\dot x)=\frac{q}{2}(\bold r \times \bold v)$. To obtain value of work and its distribution, we have solved equation (5) and (6) numerically using verlet algorithm \cite{22}. We first evolve the system upto a large time greater than the typical relaxation time so that the system is in equilibrium and then apply a time dependent protocol for the magnetic field. We have calculated values of the work for $10^5$ different realisations to get better statistics. The values of work obtained for different realisations can be viewed as random samples from the probability distributions $P(W)$. We have fixed friction coefficient, mass, charge and $k_BT$ to be unity. All the physical parameters are taken in dimensionless units.  

First we have taken magnetic field to vary linearly in time, i.e., $\bold B=B_0\frac{t}{\tau}\hat{z}$. Work distributions for both forward and backward protocols are obtained. In figure (1) we have plotted the distributions $P_f(W)$ and $P_b(-W)$,
for forward and backward protocol respectively, which are depicted in the insets of figure (1).

\begin{figure}[h]
\includegraphics[scale=0.3,angle=270]{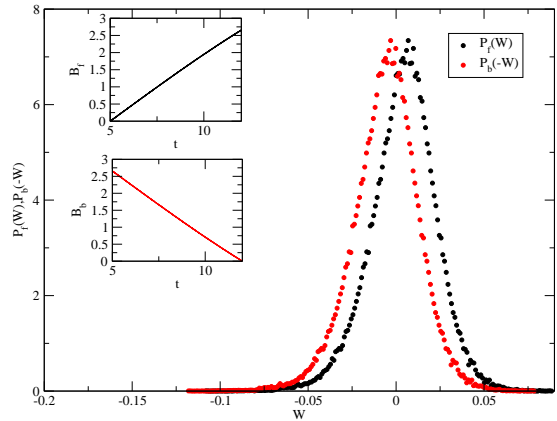}
\caption{Forward $P_f(W)$ and backward work probability distribution $P_b(-W)$ for a ramp magnetic field.}
\end{figure}
Using Jarzynski identity (equation (2))we have computed free energy difference $\Delta F$. We have obtained $\left<e^{-\beta W}\right>$ to be unity (1.0 $\pm$ 0.04) implying $\Delta F=0$. It may be noted that $\Delta F= F(B)-F(0)$, where $B$ is the value of the field at the end of the protocol. In the begining of the protocol, the value of B is zero. For different values of final magnetic field we
have obtained $\Delta F=0$ within our numerical accuracy. This implies that free energy itself (and not the free energy difference) is independent of the magnetic field, thereby satisfying the Bohr-van Leeuwen theorem as stated earlier. We can also employ Crook's equality (equation (3)) to determine the free energy difference. It follows from Crook's equality that $P_f$ and $P_b$ distributions cross at value $W=\Delta F$. This value, where the two distributions cross each other (that is, $W=0$), can be readily inferred from figure (1). This again suggests that, $\Delta F=0$ which is consistent with the result obtained using Jarzynski identity. 
\par
To strengthen our assertion (that is, the free energy being independent of magnetic field) further in figure (2) and (3) we have plotted $P_f(W)$ and $P_b(-W)$ for  two other different protocols as shown in insets of corresponding figures. For figure (3) we have considered sinusoidally varying magnetic field $B=B_1sin\omega t$ in $z$ direction. From the crossing point of $P_f$ and $P_b$ we 
observe that $\Delta F=0$, consistent with earlier result.     

\begin{figure}[h]
\includegraphics[scale=0.3,angle=270]{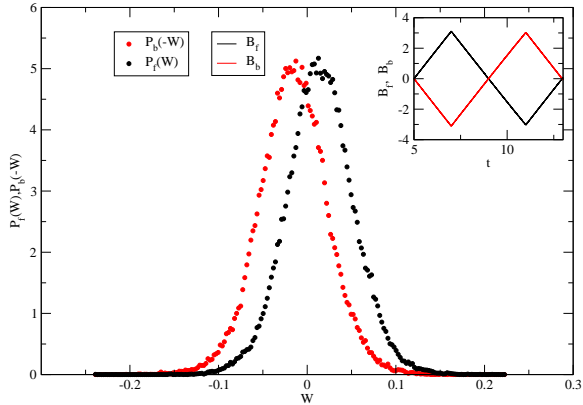}
\caption{ $P_f(W)$and $P_b(-W)$ for symmetric ramp for B(t).}
\end{figure}

\begin{figure}[h]
\includegraphics[scale=0.3,angle=270]{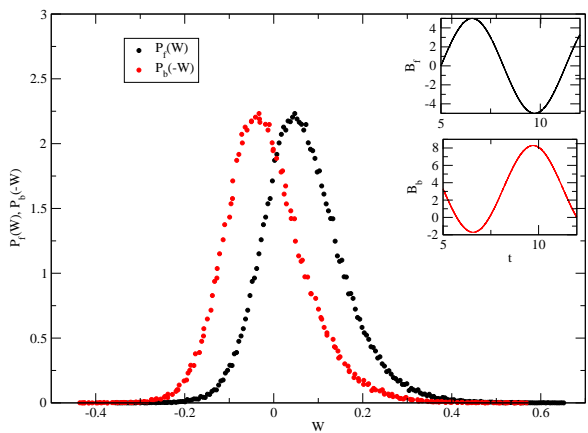}
\caption{$P_f(W)$ and $P_b(-W)$ for oscillatory magnetic field}
\end{figure}

In figure (4) we have plotted $P_f(W)$ and $P_b(-W)e^{\beta W_d}$, corresponding to the protocol shown in figure (3). Both the graphs fall on each other (within numerical error), thus verifying Crook's equality. It may be noted that reverse protocol also implies  reversing the magnetic field  \cite{23}. In all our figures the distribution of work is asymmetric and depends
on the magnetic field protocol explicitly as opposed to $ \Delta F$. Moreover, all the distributions show significant tail in
the negative work region. This is necessary  so as to satisfy Jarzynski identity. 
    
\begin{figure}[h]
\includegraphics[scale=0.32,angle=270]{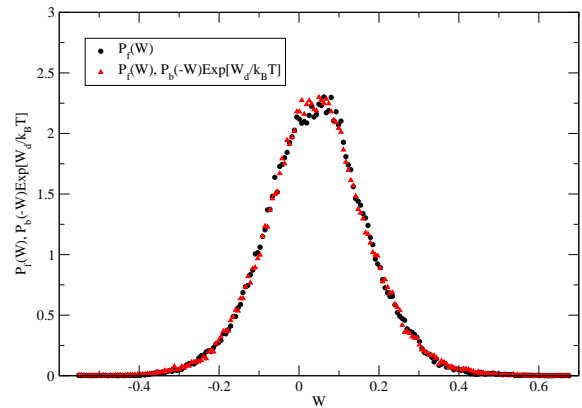}
\caption{$P_f(W)$ and $P_b(-W)e^{\beta W_d}$ are plotted together.}
\end{figure}

\begin{figure}[h]
\includegraphics[scale=0.32,angle=270]{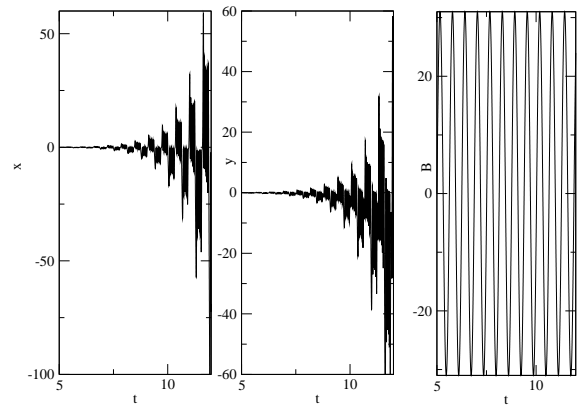}
\caption{x and y Coodinates of the particle and the $B(t)$ are plotted as a function of time.}
\end{figure}

We now discuss very briefly the occurrence of parametric resonance \cite{24} in our system in presence of sinusoidally oscillating magnetic field $B(t)=B_1sin\omega t$. In the parameter range $\frac{q_1B_1}{4\sqrt{2L_1}}-\Gamma_1>0$ where $L_1=1+\frac{2(k_1-\Gamma_1^2)}{q_1^2B_1^2}$ (see Appendix), our system exhibits instability. Here $k_1=\frac{k}{m}$, $\Gamma_1=\frac{\Gamma}{2m}$ and $q_1=\frac{q}{2m}$. The external parametric magnetic field injects energy into the system and this pumping is expected to be strongest near twice the systems frequency ($\sqrt L_1$). The trajectory of the Brownian particle grows exponentially in time also exhibiting oscillatory motion at twice the frequency of external magnetic field. This is shown in figure (5), where the coordinates of the particle and the protocol are plotted as a function of time. The parameters are $B_1=60$ and $\omega=1$. For these graphs noise strength $k_BT$ is taken as one. In presence of this instability (large variation in coordinate values) it becomes difficult to calculate work distributions as it requires large number of realisations and better accuracy. Further work in analysing the nature of the parametric resonance and associated work distributions is in progress.

In conclusion, by considering the dynamics of a trapped charged Brownian particle in a time dependent magnetic field we have verified  Jarzynski identity and Crook's equality. As a by product our result complements Bohr-van Leeuwen theorem. Work done on the system by external field arises due to the time variation of vector potential. This is in contrast to earlier studied models where the input energy to the system comes from time variation of the coordinate dependent potentials. Finally, we have discussed very briefly the occurrence of parametric resonance in our system. Our results are amenable to experimental verification. 

\section{Appendix}
\title{Occurance of parametric resonance in our system}
In presence of oscillatory magnetic field $B(t)=B_1sin\omega t$, the mean values of coordinates $<x>$, $<y>$ of the particle (averaged over thermal noise) obey the following equation for $z=<x>+i<y>$ 
\begin{equation*}
m\ddot z+(\Gamma +iqB_1sin\omega t)\dot z+(k+i\frac{qB_1\omega}{2}cos\omega t)z=0 ~~~~(A1),
\end{equation*} 
With $k=mk_1$, $\Gamma=m\Gamma^\prime$, $q=mq^\prime$ the above equation becomes
\begin{equation*}
\ddot z+(\Gamma^\prime +iq^\prime B_1sin\omega t)\dot z+(k_1+i\frac{q^\prime B_1\omega}{2}cos\omega t)z=0 ~~~~(A2).
\end{equation*}
Now, using the following transformation,
\begin{equation*}
z(t)=\xi (t)\exp [-\frac{1}{2}\int^t (\Gamma^\prime+iq^\prime B_1sin\omega t)dt], ~~~~~~~~~~~~(A3)
\end{equation*}
equation (A2) becomes
\begin{equation*}
\ddot \xi+[k_1-\frac{1}{4}(\Gamma^\prime+iq^\prime B_1sin\omega t)^2]\xi=0. ~~~~~~~~~~~~~~~~~~~~~(A4)
\end{equation*}
Redefining $\Gamma^\prime$ and $q^\prime$ as $\Gamma_1=\Gamma^\prime /2$ and $q_1=q^\prime /2$ we get,
\begin{equation*}
\ddot \xi+[k_1-(\Gamma_1+iq_1B_1sin\omega t)^2]\xi=0.~~~~~~~~~~~~~~~~~~~~~(A5)
\end{equation*}
 Again after transforming $t$ as $t=\frac{\sqrt 2 t_1}{q_1B_1} - \frac{\pi}{2\omega}$ and $\omega$ as $\omega=\frac{\omega_1q_1B_1}{\sqrt2}$ we get,
\begin{equation*}
\frac{d^2\xi}{dt_1^2}+[L_1+cos2\omega_1 t_1+i\epsilon cos\omega_1 t_1]\xi=0, ~~~~~~~~~~~~~~~~~~(A6)
\end{equation*}
 where, $L_1=1+\frac{2(k_1-\Gamma_1^2)}{q_1^2B_1^2}$ and $\epsilon=\frac{4\Gamma_1}{q_1B_1}$.
 For large $B_1$, $\epsilon$ is a small parameter and hence $i\epsilon cos\omega_1 t_1$ can be treated as perturbative term as long as $\omega_1$ is far from $2\sqrt L_1$. The condition $L_1>1$ should be maintained. Thus $\xi$ can be expanded as $\xi=\xi_0+\epsilon \xi_1+...$. Using this in equation(A6), we get (keeping only $\epsilon^0$ order term), 
\begin{equation*}
\frac{d^2\xi_0}{dt_1^2}+[L_1+cos2\omega_1 t_1]\xi_0=0.~~~~~~~~~~~~~~~~~~~~~~~~~~~~(A7)
\end{equation*} 
The above equation exhibits parametric resonance \cite{24}when $\omega_1\approx \sqrt L_1$.
Near resonating frequency, $\xi_0$ goes as $\xi_0\sim e^{st_1}$, where $s\approx \frac{1}{4\sqrt L_1}$.
Hence, $z$ will grow exponentially, if $st_1-\Gamma_1t>0$, i.e., $\frac{q_1B_1}{4\sqrt{2L_1}}(t+\frac{\pi}{2\omega})-\Gamma_1 t>0$.The condition given above can be maintained if $\frac{q_1B_1}{4\sqrt{2L_1}}-\Gamma_1 \geq 0$. For small amplitude of magnetic field, the trajectories of the particles is stable. \\ \\

\end{document}